\documentclass[reprint,aps,pra,groupedaddress,nofootinbib,floatfix,twocolumn,showpacs]{revtex4-1}
\usepackage{dcolumn}
\usepackage{bm}
\usepackage{cancel}
\usepackage{amsfonts}
\usepackage{amsmath}
\usepackage{amssymb}
\usepackage{graphicx}

\begin{document}

\title{Universal trapping law induced by atomic cloud in single-photon cooperative dynamics}
\author{Lei Qiao$^{1}$ and Chang-Pu Sun$^{1,2}$}
\email{cpsun@csrc.ac.cn}
\date{July 29, 2019}
\address{$^{1}$Graduate School of China Academy of Engineering Physics,
Beijing 100193, China} \address{$^{2}$Beijing Computational Science Research
Center, Beijing 100193, China}

\begin{abstract}
Single-photon cooperative dynamics of an assembly of two-level quantum
emitters coupled by a bosonic bath are investigated. The bosonic bath is
general and it can be anything as long as the exchange of excitations between
quantum emitters and bath is present. In these systems, it is found that the
population on the excited emitter keeps a simple and universal trapping law
due to the existence of system's dark states. Different from the trapping
regime caused by photon-emitter dressed states, this type of trapping is only
associated with the number of quantum emitters. According to the trapping law,
the cooperative spontaneous emission at single-photon level in this kind of
systems is universally inhibited when the emitter number is large enough.
\end{abstract}

\maketitle

\vspace*{3.1mm}

Cooperative light-matter interaction plays an important role in quantum
electrodynamics \cite{Dicke54,Mandel95} and is useful for various applications
of quantum optics such as optical quantum-state storage
\cite{Eisaman04,Kalachev06,Kalachev07}, quantum communication
\cite{Kuzmich03,Eisaman05}, and quantum information processing \cite{Porras08}%
. For a single excitation of an ensemble of quantum emitters, the rate and the
direction of the cooperative spontaneous emission can be strongly modified by
different light-field environments. While the size and the shape of the
ensemble have been investigated \cite{Scully06,Li06,Miroshnychenko13}, an
ensemble of atoms with a single collective excitation also exhibits a dynamics
characterized by revivals for different atom numbers in a bosonic bath with
linear dispersion relation \cite{Kumlin18}.

However, until now, almost all the results and conclusions about the
single-photon cooperative dynamics provided by the published papers are based
on the specific light-field environments and the specific coupling coefficient
between quantum emitters and photon \cite%
{Cummings83,FWC85,Benivegna88,Buzek89,Buzek99,AAS08,SD08,Scully09,AAS10,SZ12,Li12,Boag13,Xu13,Feng14,Liao15,Jenkins17}. If the light-field environments and
coupling coefficient are changed, will these results and conclusions change or
keep the same?

Here, we focus on the single-photon cooperative dynamics in a system that
the light-field environment and the coupling coefficient are general and
physical. The emitters are assumed to be placed much closer than the
wavelength of radiation field and thus the emitters are efficiently coupled by
the radiation field without retardation effects.

In this paper, we report that there is a universal trapping law in the
single-photon cooperative dynamics based on an analytical analysis which is
beyond Wigner-Weisskopf approximation and Markovian approximation. A direct
conclusion comes from this law is that the spontaneous emission dynamics in
this system is suppressed if the number of the emitters is large enough.

We begin with the system that contains $M$ two-level atoms coupled to the
radiation field in an environment with a general dispersion relation
$\omega_{k}$. The atoms are characterized by ground state $\left\vert
g\right\rangle $ and excited state $\left\vert e\right\rangle $. The
Hamiltonian of this system in the rotating-wave approximation takes the form
(with $\hbar=1$)%
\begin{align}
H  & =\sum_{k}\omega_{k}a_{k}^{\dag}a_{k}+\sum_{j=1}^{M}\Omega_{j}\left\vert
e_{j}\right\rangle \left\langle e_{j}\right\vert \nonumber\\
& +\sum_{j,k}V_{k,j}\left(  \sigma_{j}^{+}a_{k}+\sigma_{j}^{-}a_{k}^{\dag
}\right)
\end{align}
where the first term describes the light field and $a_{k}^{\dag}$ ($a_{k}$)
denotes the creation (annihilation) operator of photon with momentum $k$. The
second term represents two-level atoms and $\Omega_{j}$ is atom's transition
frequency. Here, we set the ground energy of atoms to be zero as reference.
The last term represents the interaction between photon and atoms. $\sigma
_{j}^{+}=\left\vert e_{j}\right\rangle \left\langle g_{j}\right\vert $
($\sigma_{j}^{-}=\left\vert g_{j}\right\rangle \left\langle e_{j}\right\vert
$) is the raising (lowering) operator acting onto the $j$th atom and $V_{k,j}$
is the coupling strength.

To investigate the dynamics of atoms when one of them is excited, we start
from the time-dependent Schr\"{o}dinger equation%
\begin{equation}
i\frac{\partial}{\partial t}\left\vert \psi\left(  t\right)  \right\rangle
=H\left\vert \psi\left(  t\right)  \right\rangle \label{TSchEq}%
\end{equation}
where $|\psi\left(  t\right)  \rangle$ is the state of the system at time $t$.
Since the total excitation number $N=\sum_{k}\omega_{k}a_{k}^{\dag}a_{k}%
+\sum_{j=1}^{M}|e_{j}\rangle\langle e_{j}|$ is conserved, the state
$|\psi\left(  t\right)  \rangle$ with $N=1$ can be expanded as $|\psi
(t)\rangle=\sum_{j}A_{j}(t)|g_{1}g_{2}...e_{j}...g_{M},0\rangle+\sum_{k}%
C_{k}(t)|g_{1}g_{2}...g_{M},1_{k}\rangle$, where $A_{j}(t)$ is the probability
amplitude of the state with $j$th atom in the excited state and the others in
the ground states and no photon in the environment, while $C_{k}(t)$
represents the probability amplitude for finding all atoms to be in the ground
states and one photon in the environment. Take $|\psi(t)\rangle$ into Eq.
(\ref{TSchEq}), one obtains the equations for $A_{j}(t) $ and $C_{k}(t)$%
\begin{equation}
i\frac{\partial A_{j}\left(  t\right)  }{\partial t}=\Omega_{j}A_{j}\left(
t\right)  +\sum_{k}V_{k,j}C_{k}\left(  t\right)  \text{,}\label{At}%
\end{equation}%
\begin{equation}
i\frac{\partial C_{k}\left(  t\right)  }{\partial t}=\omega_{k}C_{k}\left(
t\right)  +\sum_{j}V_{k,j}A_{j}\left(  t\right)  \text{.}\label{Ct}%
\end{equation}
The method based on Wigner-Weisskopf approximation or Markovian approximation
theory is widely used to solve the dynamical equations in Eq. (\ref{At}) and
(\ref{Ct}) with specific $\omega_{k}$ and $V_{k,j}$, which leads to a result
that the excited atomic population reveals exponential decay or the population
decay is complete. However, it has been pointed out that an important
information about the population trapping will be lost when one of this two
kinds of approximation theories is used \cite{John94,Tudela17,Tudela172}.

To go beyond Wigner-Weisskopf approximation and Markovian approximation, we
take a Laplace transform of Eq. (\ref{At}) and (\ref{Ct}) and it gives%
\begin{equation}
i\left[  -A_{j}\left(  0\right)  +s\tilde{A}_{j}\left(  s\right)  \right]
=\Omega_{j}\tilde{A}_{j}\left(  s\right)  +\sum_{k}V_{k,j}\tilde{C}_{k}\left(
s\right)  \text{,}\label{LAt}%
\end{equation}%
\begin{equation}
i\left[  C_{k}\left(  0\right)  +s\tilde{C}_{k}\left(  s\right)  \right]
=\omega_{k}\tilde{C}_{k}\left(  s\right)  +\sum_{j}V_{k,j}\tilde{A}_{j}\left(
s\right)  \text{.}\label{LCt}%
\end{equation}
Denote the initial excited atom as $j_{0}$, i.e., the initial amplitudes are
$A_{j_{0}}\left(  0\right)  =1$, $A_{j}\left(  0\right)  =0$ ($j\neq j_{0}$)
and $C_{k}\left(  0\right)  =0$, the expression of $\tilde{A}_{j_{0}}\left(
s\right)  $ can be acquired
\begin{equation}
\tilde{A}_{j_{0}}\left(  s\right)  =i\frac{is-\Omega-(M-1)f\left(  s\right)
}{\left(  is-\Omega\right)  \left[  is-\Omega-Mf\left(  s\right)  \right]  }%
\end{equation}
where $f\left(  s\right)  \equiv\sum_{k}V_{k}^{2}/(is-\omega_{k})$. Here it
has been assumed that the atoms are identical and thus $\Omega_{1}=\Omega
_{2}=...=\Omega_{M}=\Omega$, $V_{k,1}=V_{k,2}=...=V_{k,M}=V_{k}$. The
amplitude $A_{j_{0}}\left(  t\right)  $ is given by the inverse Laplace
transform $A_{j_{0}}\left(  t\right)  =\frac{1}{2\pi i}\int_{\sigma-i\infty
}^{\sigma+i\infty}\tilde{A}_{j_{0}}\left(  s\right)  e^{st}ds$, which leads to%
\begin{align}
A_{j_{0}}\left(  t\right)   & =\sum_{n}\frac{s+i\Omega+i(M-1)f\left(
s\right)  }{[F\left(  s\right)  ]^{\prime}}e^{st}|_{s=x_{n}^{\left(  1\right)
}}\nonumber\\
& -\int_{C}\frac{s+i\Omega+i(M-1)f\left(  s\right)  }{2\pi iF\left(  s\right)
}e^{st}ds\label{Aj0t}%
\end{align}
where $F\left(  s\right)  \equiv(s+i\Omega)[s+i\Omega+iMf\left(  s\right)  ]$
and $[F\left(  s\right)  ]^{\prime}$ means the derivative of $F\left(
s\right)  $ with respect to $s$. $x_{n}^{\left(  1\right)  }$ is the roots of
the equation $F\left(  s\right)  =0$ in the complex plane except the regions
in order to ensure that the integrand is single-valued function. $C$ is the
integration contour based on residue theorem. Generally, $C$ is associated
with the specific expression of $V_{k}$ and $\omega_{k}$. Different $V_{k}$
and $\omega_{k}$ lead to different integration contour $C$. However, a common
conclusion that does not depend on specific $V_{k}$ and $\omega_{k}$ is that
the second term of $A_{j_{0}}\left(  t\right)  $ in Eq. (\ref{Aj0t}) goes to
zero when time $t$ tends to infinity due to the factor $e^{st}$ \cite{Riley06}.

\begin{figure}[ptb]
\begin{center}
\includegraphics[width=7.5cm]{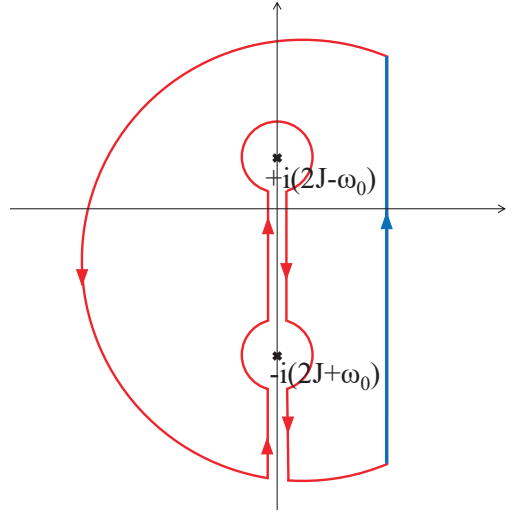}
\end{center}
\caption{Integration contours for the calculation of $A_{j_{0}}(t)$ in the
coupled-cavity system. The red line is the integration contour $C$.}
\label{fig1}
\end{figure}
\begin{figure}[ptb]
\begin{center}
\includegraphics[width=8.2cm]{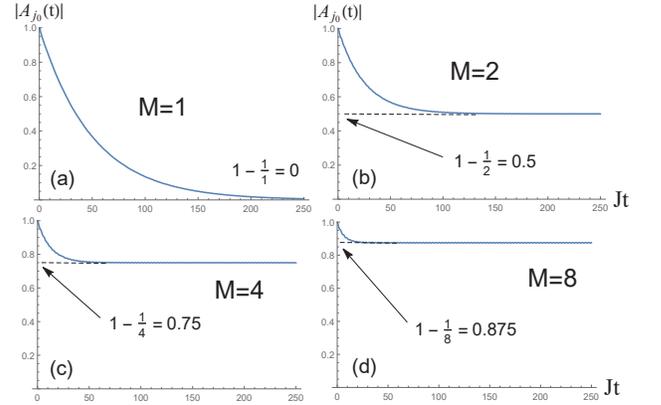}
\end{center}
\caption{Time evolution of the population $|A_{j_{0}}\left(  t\right)  |$ on
the excited atom with different atom number in the coupled-cavity system. The
coupling strength $g_{0}=0.2J$. The detuning $\delta_{1}\equiv\Omega
-\omega_{0}=0$.}
\label{fig2}
\end{figure}

The physics in Eq. (\ref{Aj0t}) is not obvious. We transform Eq. (\ref{Aj0t})
into another form which is the key point for the analysis in the following
\begin{align}
A_{j_{0}}\left(  t\right)   & =\frac{M-1}{M}e^{-i\Omega t}+\sum_{m}%
\frac{e^{st}}{M\left[  G\left(  s\right)  \right]  ^{\prime}}|_{s=x_{m}%
^{\left(  2\right)  }}\nonumber\\
& -\int_{C}\frac{G(s)-if\left(  s\right)  }{2\pi iF\left(  s\right)  }%
e^{st}ds\label{Aj0t1}%
\end{align}
where $G(s)\equiv s+i\Omega+iMf\left(  s\right)  $ and $x_{m}^{\left(
2\right)  }$ is the roots of the equation $G(s)=0$. Here, the equation
$G(-iE)=0$ is nothing but the system's eigenenergy equation of photon--atom
dressed state. In fact, the second term of $A_{j_{0}}\left(  t\right)  $ in
Eq. (\ref{Aj0t1}) comes from system's photon-atom bound states that the
populations of field modes are not zero and the third term comes from system's
scattering states \cite{John94,Qiao19}. The first term in Eq. (\ref{Aj0t1}) is
only related with the atom's transition frequency and the number of atoms. It
comes from system's dark state with energy $\Omega$ that all the excitation number focuses on the atoms and the
populations of field modes are zero \cite{Sun03,S2un03}. This kind of dark
state is universal in this kind of system. It is caused by the collective
coherence of atomic clouds. So the trapping associated with the dark state is
universal no matter whether the role of the second term in Eq. (\ref{Aj0t1})
is important or not.

\begin{figure}[tbp]
\begin{center}
\includegraphics[width=7.5cm]{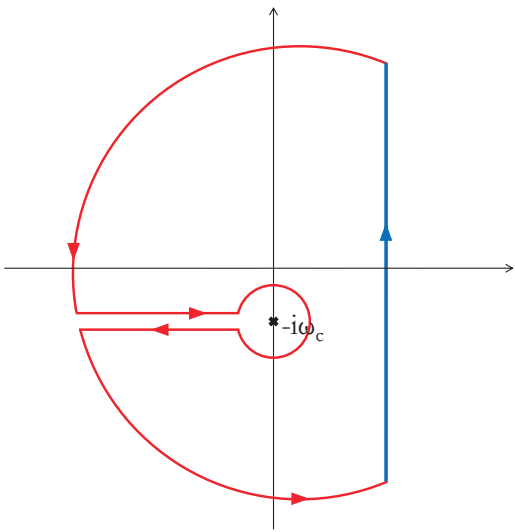}
\end{center}
\caption{Integration contours for the calculation of $A_{j_{0}}(t)$ in the
photonic crystal system. The red line is the integration contour $C$.}
\label{fig3}
\end{figure}
\begin{figure}[tbp]
\begin{center}
\includegraphics[width=8.4cm]{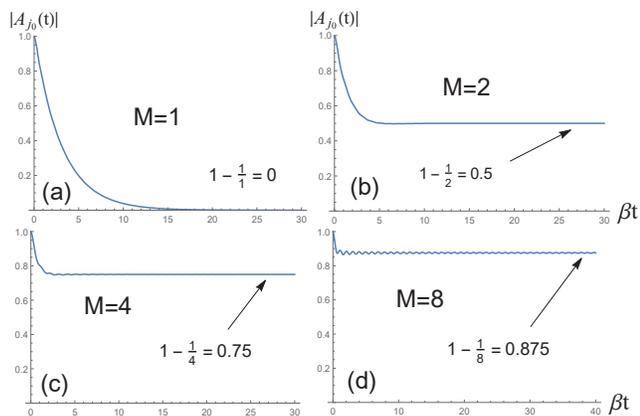}
\end{center}
\caption{Time evolution of the population $|A_{j_{0}}\left(  t\right)  |$ on
the excited atom with different atom number in the photonic crystal system.
The detuning $\delta_{2}\equiv\Omega-\omega_{c}=6.5\beta$. Here $\beta
^{3/2}\equiv(\Omega^{2}d^{2})/(6\pi\epsilon_{0}B^{3/2})$.}
\label{fig4}
\end{figure}

When the equation $G(s)=0$ has no roots or that the system's parameters
satisfy the condition $|1/\{M[G(x_{m}^{(2)})]^{\prime}\}|<<1$, the final
result of the amplitude $A_{j_{0}}\left(  t\right)  $ at $t=\infty$ is%
\begin{equation}
\left\vert A_{j_{0}}\left(  \infty\right)  \right\vert =1-\frac{1}%
{M}\label{UniTr}%
\end{equation}
which is only related with the atom number. This trapping phenomenon takes
place when the number $M>1$.

To check this universal trapping, we now present two examples. One is the
system of one-dimensional coupled-cavity waveguide, in which the dispersion
$\omega_{k}=\omega_{0}-2J\cos(k)$ and the coupling coefficient $V_{k}=g_{0}$
\cite{Zhou08,Longo10,Zhou13}. Here $\omega_{0}$ is the on-site energy of each
cavity and $J$ represents the hopping energy of the photon between two
neighbouring cavity. The other is the system of three-dimensional photonic
crystal with $\omega_{\mathbf{k}}=\omega_{c}+B(\mathbf{k}-\mathbf{k}_{0})^{2}$
and $V_{\mathbf{k}}=\Omega d\sqrt{\frac{1}{2\epsilon_{0}\omega_{\mathbf{k}}V}%
}\mathbf{e}_{k}\cdot\mathbf{u}$ \cite{John90,John94,Zhu97,Lambropoulos00}. $d$
and $\mathbf{u}$ are the magnitude and unit vector of atomic dipole moment.
$V$ is the volume and $\mathbf{e}_{k}$ is the two transverse unit vectors of
polarization. Both of the two systems have been extensively studied
theoretically and experimentally in recent years. For the coupled-cavity
system, the integration contour $C$ is shown by the red line in Fig. 1. When
$\Omega=\omega_{0}$, the condition $|1/\{M[G(x_{m}^{(2)})]^{\prime}\}|<<1$ can
be easily satisfied. In Fig. 2, we plot the time evolution of $A_{j_{0}%
}\left(  t\right)  $ for different number of atoms. One see that $|A_{j_{0}%
}(\infty)|$ meets the value $1-1/M$. For the photonic crystal system, the
integration contour $C$ is plotted with the red line in Fig. 3. The time
evolution of $A_{j_{0}}\left(  t\right)  $ is shown in Fig. 4. The trapping
law $1-1/M$ is also obeyed when the condition $|1/\{M[G(x_{m}^{(2)})]^{\prime
}\}|\ll1$ is satisfied.

To sum up, we have explored the single-photon cooperative dynamics in an
ensemble of two-level atoms which is couple to a general bosonic bath. The
size of the ensemble is much smaller than the wavelength of radiation field.
The bosonic bath can be photonic crystal, waveguide, or anything else as long
as the exchange of excitations between atoms and bath can take place. It is
found that there is a universal trapping caused by system's dark state. This
kind of trapping obeys a simple law that is only related with the number of
atoms. A direct conclusion comes from this law is that the single-photon
cooperative spontaneous emission is suppressed when there are many enough
atoms. Besides, due to the presence of this trapping, the energy of the
radiation field will be less than the initial total energy $E_{tot}=\Omega$.

\section{Acknowledgements}

This work is supported by the National Basic Research Program of China
(Grant No. 2016YFA0301201 \& No. 2014CB921403), the NSFC (Grant No.
11534002), and the NSAF (Grant No. U1730449 \& No. U1530401).

\end{document}